# Business Process Modeling Notation - An Overview

**Assist.Prof. Alexandra Fortiş, PhD Candidate**
"Tibiscus" University, Timişoara, România

**REZUMAT.** BPMN reprezintă un standard industrial creat pentru a oferi o notaţie comună şi uşor de înţeles pentru toţi participanţii la un proces de afaceri. În această lucrare propunem o prezentare succintă a principalelor caracteristici ale acestei notaţii precum şi o interpretare a unor dintre principalele şabloane care caracterizează un proces de afaceri modelat prin intermediul fluxurilor de lucru.

## 1 Introduction

Business Process Modeling Notation (BPMN) represents a new industrial standard created in order to provide a graphical notation accessible to business and technical users involved in a business process modeled through a workflow. This graphical notation was developed by the Business Process Management Initiative (BPMI) and one of its main goals was to mark and to coordinate the sequence of processes and the change of messages between the participants involved in a business process.

BPMN was created in order to promote and develop the use of Business Process Management (BPM) through the establishment of open standards for process design, deployment, execution, maintenance, and optimization.

Through BPMN, a complex business process can be easily coordinated despite the large number of participants. The benefit consists in the fact that we have a free notation readily understandable by all process users, from business analysts to technical developers. Thus has been covered the gap between the process design and the process implementation. Another important achievement is the interconnectivity between BPMN and XML languages designed for the execution of business processes, such as BPEL4WS (Business Process Execution Language for Web Services) and BPML (Business Process Modeling Language), can be visually expressed with a common notation.

BPMN can be used to model *internal business processes*, named BPM processes, defining activities that are not visible to the public, *abstract (public) processes* showing activities used only for communication and ignoring all other internal activities and *collaboration (global) processes*, marking the sequence of all activities involved and the exchange of messages between the actors.

Except BPMN, BPMI has designed BPML (Business Process Modeling Language) as being the standard execution language and proposed BPQL (Business Process Query Language) for a standard management interface.

## 2  Basic elements of BPMN

BPMN is a π-Calculus based standard for the description of business processes. BPMN uses a *Business Process Diagram* (BPD) for expressing business processes. This diagram was designed for being easily read and used, to offer expressiveness to very complex business processes models and to be rapidly mapped to business execution languages.

To create a model for a business process we have to describe the natural course of the process: what starts the process, what processes are performed and their result.

There are three basic model types that can be created by using a BPD:
- *Collaboration* business processes (B2B), showing the interactions between two or more entities. These give the choreographic part of the language;
- *Private* business processes, focused on the activities of a single business organization (not visible to the public – private)
- *Abstract* business processes, emphasizing activities used by a business process in order to communicate with another process or participant.

In a BPD, decisions are modeled through *gateways* and a process may be modeled using another BPD. If the process is not composed from other sub-processes, it is called *task* and it represents the lowest level of the process. The sign "+" is used to denote the fact that a process is decomposed into sub-processes. Its absence is characteristic to a task. We use pools partitioned into lanes in order to group events and processes.

## 2.1 Main concepts

The modeling in BPMN is made by simple diagrams with a small set of graphical elements. It should make it easy for business user as well as developers to understand the flow and the process. The basic categories of elements are:
- Flow Objects, like events, activities, gateways;
- Connecting Objects, like *sequence flow*, *message flow*, *association*;
- Swimlanes, like *pool*, *lane*;
- Artifacts, like *data objects*, *text annotations*, *groups*.

Models can be produced with a higher level of detail, by using supplemental markers for the core events, like the markers for Message, Time, Exception, Compensation, and others (see Figure 1). Also, one can add markers to *activities* in order to make a distinction between *Loop*, *Multiple Instance*, *Compensation*, or *Ad-Hoc* activities. Thus, one can use BPMN to create high-level processes as well as complex processes with multiple levels of detail.

## 2.2 Error handling, transactions, and compensation support

BPN supports *exception handling*. Thus, intermediate events that are attached to the boundary of an activity are triggers that can interrupt the activity.

*Transactions* are activities drawn with doubled borders. One can model various kinds of flows, specifying a successful completion of the transaction, a cancelled completion of the transaction, or an exception causing the failure of a transaction.

BPMN offers support for *compensation*. Thus, one can designate some activities as being able to roll-back the effects of another activity. Notice that activities that are used with the compensation marker are outside the normal flow of the process, and are considered to be *associated normal activities*.

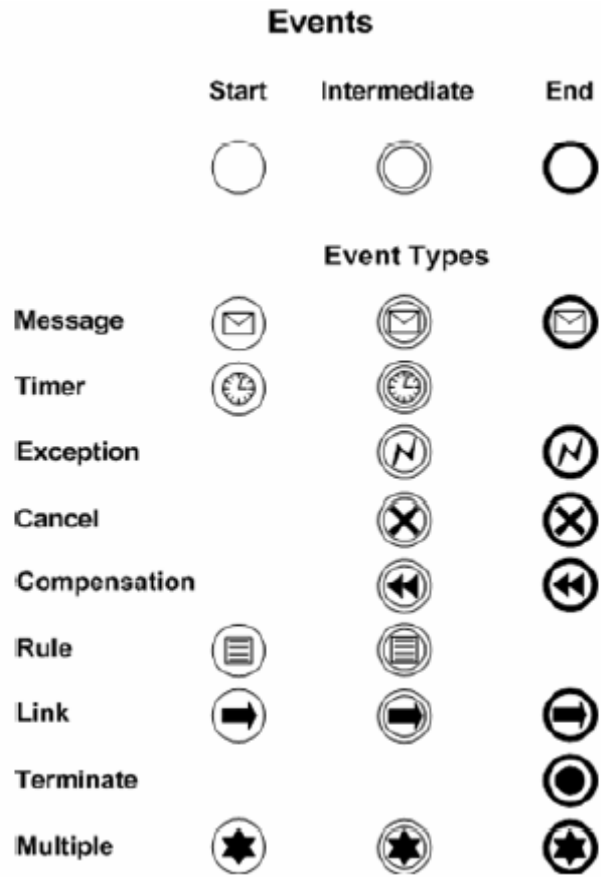

***Fig. 1:*** *Types of markers that can be added to BPMN core events [Whi05b]*

## 3   BPMN and workflow patterns

In their paper [Aal03] van der Aalst and his team, presented a set of patterns on which they built the analysis of workflows. White [Whi04a] refined their research and extended it to BPMN, considering a *Normal Sequence Flow* that includes a *Forking Flow, Joining Flow, Splitting Flow, Merging Flow*, and *Looping*, as well as an *Exception Flow*.

As mentioned above, we can model decisions, merges, forks, and joins in the process flow using a **gateway** symbol. At this point, we expect a question to be answered in multiple ways. We shall present in this paper some of possible interpretations for a gateway with graphical explanation for advanced patterns for workflow, as defined by Workflow Management System.

For the **Exclusive choice pattern** we can have two possible situations: data-based XOR decision and event-based XOR decision. For both of them, the gateway represents the point in the process where, a decision is made by choosing one of the possible paths. The decision is marked by a specific symbol for BPMN, named exclusive decision gateway.

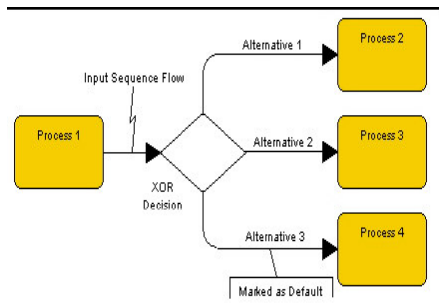 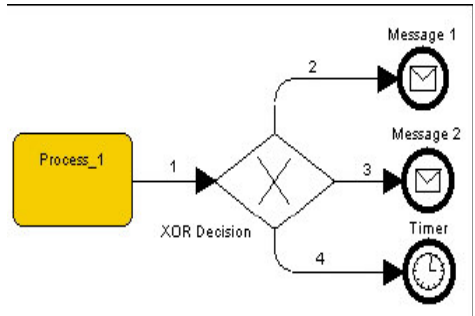

*Fig. 2: Data-based XOR decision*     *Fig. 3: Event-based XOR decision*

The **exclusive merge pattern** represents a point in a workflow where two or more branches are joined without synchronization and none of these branches is executed in parallel with another one. For this pattern BPMN uses the same **exclusive decision gateway** as mentioned before but is possible to model the pattern directly. In that case, the joining operation can not be controlled.

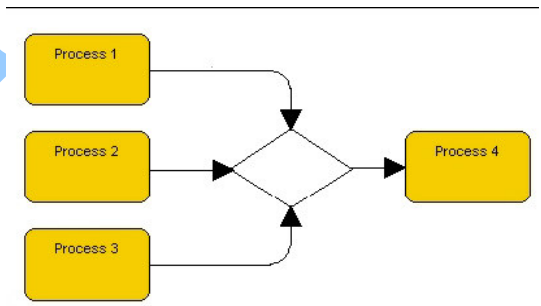

*Fig. 4: Exclusive XOR merge*

45

The **multiple choice pattern** marks a point in a workflow, as defined by Workflow Management System, where, after considering a certain decision, a number of branches can be selected.

This pattern is used in order to model conditioned consecutive tasks. To represent it, BPMN uses an OR-split operator, named **inclusive decision gateway**.

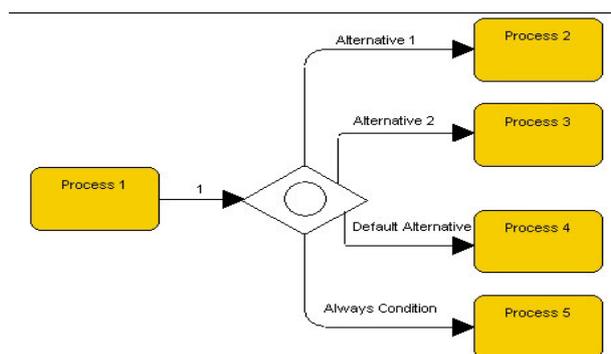

***Fig. 5:*** *Inclusive OR*

The **synchronizing merge pattern** marks a point in a workflow, as defined by Workflow Management System where multiple paths converge to a single output under the condition of synchronization, meaning that once activated, a path can not be reactivated while the process waits for the completion of other paths. The major problem in this situation consists in the fact that the decision to be made can be either synchronization or a joining operation. BPMN uses an OR-join operator marking the corresponding gateway.

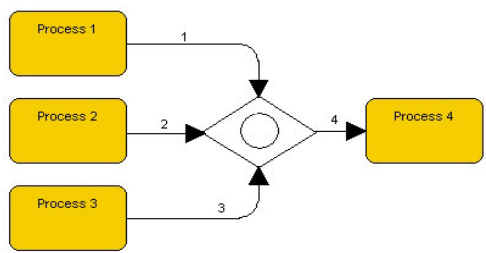

***Fig. 6:*** *Inclusive OR merge*

In the case of a complex decision, the modeler has to specify by a complex condition, which output flow will be followed. The pattern modeled here is **parallel split**, marking the point of a workflow, as defined by Workflow Management System where a single path splits into multiple branches that can be executed in parallel.

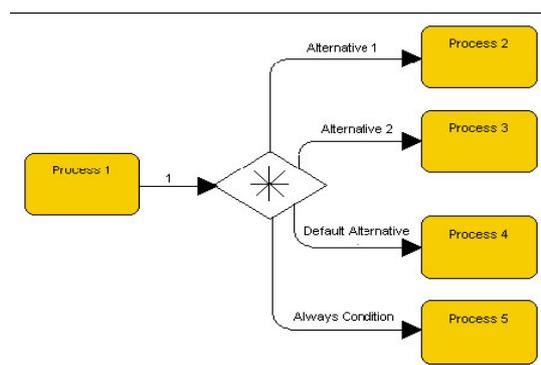

***Fig. 7:*** *Complex decision*

The complex merge is a situation characterizing the moment when a certain task will begin.

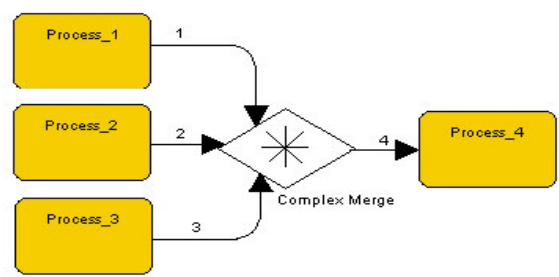

***Fig. 8:*** *Complex merge*

Another two workflow patterns are presented: **parallel forking** (all of the output paths are followed) and **parallel joining** (the process flow expects all of the input signals to arrive at the gateway in order to obtain an output answer).

47

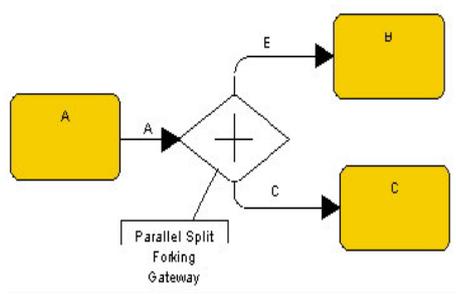 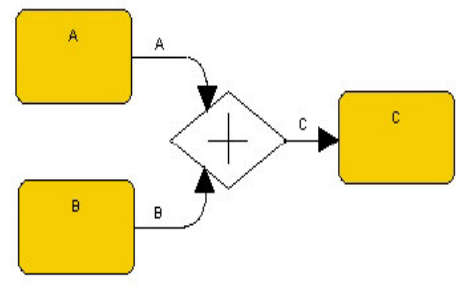

*Fig. 9: Parallel forking*      *Fig. 10: Parallel joining*

**Conclusions**

One of the major advantages of BPMN is that it is extensible, meaning that the diagram elements can be modified by modelers by adding graphical elements that could bring more details to the diagrams for an easier understanding.

BPMN was created to provide a common language for modeler analysts and developers by mapping to an executable language and by the support for B2B process concepts, public and private. It also enables exception handling, transactions and compensation to be modeled.

BPMN can model the exchange of messages between the actors of a business process and provides a notation for data objects that can be associated to activities.